\documentstyle[12pt,fleqn,epsf]{article}
\newcommand{\bea}{\begin{eqnarray}}
\newcommand{\beq}{\begin{equation}}
\newcommand{\eea}{\end{eqnarray}}
\newcommand{\eeq}{\end{equation}}
\newcommand{\nnu}{\nonumber}

\def\eq#1{{eq.~(\ref{#1})}}

\def\Tr{\mathop{\mbox{Tr}}\,}

\begin{document}
\title{Feynman rules for the strong and $\Delta S=1$ weak chiral Lagrangian.}
{\small
\author{ E.I. Lashin \\
Abdus Salam International Center for Theoretical Physics, Trieste,
Italy. \\
\& \\
Ain Shams University, Faculty of Science,\\
 Dept. of Physics,
Cairo, Egypt.} } \maketitle
\begin{abstract}
We present  Feynman rules for the strong and $\Delta S = 1$ weak
chiral Lagrangian which are relevant for calculating $K\rightarrow
\pi\pi\pi$ decay amplitudes. Feynman rules for this specific case
turn out to be too lengthy to be included in a research paper.
Since they are not standard, we decide to include them in a
separate article, making them available for any one who is
interested in this kind of calculation. Any found mistakes and
(/or) comments are warmly welcome  so that corrections can be
made.
\end{abstract}

\section{Introduction}
Feynman rules which are relevant for calculating $K\rightarrow
\pi\pi\pi$ decay amplitudes can be classified as follows:

Rules derived from the strong chiral Lagrangian, which is
\beq {\cal L}_{strong}^{(2)} = \frac{f^2}{4} \Tr (D_\mu
\Sigma^{\dagger} D^\mu \Sigma ) +  \frac{f^2}{2} B_0 \Tr ( {\cal
M}  \Sigma^{\dagger} + \Sigma {\cal M}^{\dagger} ) \label{sham}
\eeq
and
\beq \Sigma \equiv \exp \left( \frac{2i}{f} \,\Pi (x) \right)
\label{sigma} \eeq
where
\beq \Pi (x) = \frac{1}{\sqrt{2}}\, \left[
\begin{array}{ccc}
\frac{1}{\sqrt{2}}\, \pi^0 (x) + \frac{1}{\sqrt{6}}\, \eta_8 (x) & \pi^+ (x) & K^+ (x) \\
\pi^- (x) & -\frac{1}{\sqrt{2}}\, \pi^0 (x) +  \frac{1}{\sqrt{6}}\, \eta_8 (x) & K^0 (x)\\
K^- (x) & \bar{K}^0 (x) & - \frac{2}{\sqrt{6}} \, \eta_8 (x)
\end{array}
\right] \, . \label{defseg} \eeq
The scale $f$ is identified at the tree level with the  pion decay
constant $f_\pi$.

The other rules are derived from the  weak chiral Lagrangian,
which is
\bea {\cal L}^{(2)}_{\Delta S = 1}  & = &G_{\underline{8}}
(Q_{3-6}) \Tr \left( \lambda^3_2 D_\mu \Sigma^{\dag} D^\mu \Sigma
\right) +  \nnu \\
 & & \: G_{LL}^a (Q_{1,2}) \,
\Tr \left(  \lambda^3_1 \Sigma^{\dag} D_\mu \Sigma \right)
\Tr \left( \lambda^1_2 \Sigma^{\dag} D^\mu  \Sigma \right) +  \nnu \\
 & & \: G_{LL}^b (Q_{1,2})\, \Tr \left( \lambda^3_2 \Sigma^{\dag} D_\mu
\Sigma \right) \Tr \left(  \lambda^1_1 \Sigma^{\dag} D^\mu \Sigma
\right) \label{chi-lag} \, , \eea
where $\lambda^i_j$ are combinations of Gell-Mann $SU(3)$ matrices
defined by $(\lambda^i_j)_{lk} = \delta_{il}\delta_{jk}$ and
$\Sigma$ is defined in \eq{sigma}. The covariant derivatives in
eqs.~(\ref{sham}),(\ref{chi-lag}) are taken with respect to the
external gauge fields whenever they are present.

The chiral coefficients $G_{LL}^a$, $G_{LL}^b$,
$G_{\underline{8}}$. represent the contribution of the ten quark
operators relevant for $\Delta S =1 $ weak chiral Lagrangian.

In comparing with other works in literatures , it is convenient to
write the $\Delta S =1$ weak chiral Lagrangian in the following
form \bea {\cal L}^{(2)}_{\Delta S = 1} & = & g_{\underline{8}}
\Tr \left( \lambda^3_2 D_\mu \Sigma^{\dag}
D^\mu \Sigma\right) \ \nnu \\
&   & + g_{\underline{27}} \left[ \Tr \left( \lambda^3_2
\Sigma^{\dag} D_\mu \Sigma \right) \Tr \left( \lambda^1_1
\Sigma^{\dag} D^\mu  \Sigma \right) + \frac{2}{3} \Tr \left(
\lambda^3_1 \Sigma^{\dag} D_\mu \Sigma \right)
\Tr \left(  \lambda^1_2 \Sigma^{\dag} D^\mu \Sigma \right) \right] \ , \nnu \\
\label{chiwyler} \, \eea which involves two couplings $g_8$ and
$g_{27}$ representing respectively the octet $ (8_L, 1_R)$ and the
twenty--seven $(27_L,1_R)$ couplings, where $g_8$ and $g_{27}$ can
be written in terms of $G_{LL}^a$, $G_{LL}^b$,
$G_{\underline{8}}$.

As a final comment, our convention for Feynman rules is that all
particles  are entering the vertex. The same rules hold for the
conjugated couplings.

\section{Feynman Rules for the Strong Chiral Lagrangian}

\def\p0#1            { \pi^0(p_#1)}
\def\pp#1            { \pi^+(p_#1)}
\def\pm#1            { \pi^-(p_#1)}
\def\k0#1            { K^0(p_#1)}
\def\kp#1            { K^+(p_#1)}
\def\km#1            { K^-(p_#1)}
\def\bk#1            { \bar{K}^0(p_#1)}
\def\et#1            { \eta(p_#1)}

\def\P0            { \pi^0}
\def\Pp            { \pi^+}
\def\Pm            { \pi^-}
\def\K0            { K^0}
\def\Kp            { K^+}
\def\Km            { K^-}
\def\Bk            { \bar{K}^0}
\def\Et            { \eta}

\def\facm#1#2          { i\,\frac{#1}{#2 f^2} }

\def\facd#1#2            {\frac{#1} {#2 f^2}}
\def\fact#1#2            {\frac{#1} {#2 f^3}}
\def\facq#1#2            {\frac{#1} {#2 f^4}}
\def\facc#1#2            {\frac{#1} {#2 f^5}}
\def\facs#1#2            {\frac{#1} {#2 f^6}}

\def \sca#1#2          {p_{#1} \cdot p_{#2}}
\def \Scat#1#2#3        {p_{#1} \cdot (p_{#2}+p_{#3})}
\def \Scaq#1#2#3#4        {p_{#1} \cdot (p_{#2}+p_{#3} + p_{#4})}
\def \Scadd#1#2#3#4        {(p_{#1} + p_{#2}) \cdot (p_{#3}+p_{#4})}
\def \Scac#1#2#3#4#5       { p_{#1} \cdot ( p_{#2} + p_{#3}+ p_{#4} + p_{#5} ) }
\def \Scadt#1#2#3#4#5       { ( p_{#1} + p_{#2} ) \cdot  ( p_{#3}+ p_{#4} + p_{#5} ) }
\def \Scas#1#2#3#4#5#6      { p_{#1} \cdot ( p_{#2} +  p_{#3}+ p_{#4} + p_{#5} + p_{#6} ) }

In writing Feynman rule for the strong interaction, it is convenient to define
the following functions

{\footnotesize
\bea
A(p_1,  p_2, p_3, p_4) & = &  \frac{1}{2} (p_1 + p_2) \cdot (p_3 + p_4) - p_1 \cdot p_2 - p_4 \cdot p_3 \  \nnu \\
B(p_1,  p_2, p_3, p_4) & = &  \frac{1}{2} (p_1 + p_4) \cdot (p_2 + p_3) - p_1 \cdot p_4 - p_2 \cdot p_3 \ \nnu \\
C(p_1,  p_2, p_3, p_4) & = & (p_4 - p_3) \cdot (p_2 - p_1)  \ \nnu \\
D(p_1,  p_2, p_3, p_4) & = &  \frac{1}{2} (p_1 + p_3) \cdot (p_2 + p_4) - p_1 \cdot p_3 - p_2 \cdot p_4 . \ \nnu
\eea
}

The relevant strong vertices are:
{\footnotesize
\bea
\pm1  \pp2    \pp3  \pm4 &\quad & \facd{i}{6} B \ \nnu \\
\km1  \kp2    \pp3  \pm4 &\quad & \facd{i}{3} B \ \nnu \\
\pm1  \pp2    \p03  \p04 &\quad & - \facd{i}{3} A \ \nnu \\
\km1  \kp2    \p03  \p04  &\quad & - \facd{i}{12} A \  \nnu \\
\bk1  \k02    \pp3  \pm4  &\quad & \facd{i}{3} D   \ \nnu \\
\bk1  \k02    \p03  \p04 &\quad &  - \facd{i}{12} A \ \nnu \\
\km1  \k02    \p03  \pp4  &\quad & \facd{i}{2 \, \sqrt{2}} C  \ \nnu \\
\bk1  \k02    \et3  \p04  &\quad & \facd{i}{2\sqrt{3}} A \ \nnu \\
\km1  \k02    \et3  \pp4  &\quad & - \facd{i}{\sqrt{6}} A \ \nnu \\
\bk1  \k02    \et3  \et4  &\quad & - \facd{i}{4} A \ \nnu \\
\km1  \kp2    \et3  \et4  &\quad &  - \facd{i}{4} A \ \nnu \\
\bk1  \k02    \k03  \bk4 &\quad & \facd{i}{6} B \ \nnu \\
\km1  \kp2    \kp3  \km4 &\quad & \facd{i}{6} B \ \nnu \\
\bk1  \k02    \kp3  \km4  &\quad &  \facd{i}{3} B \ \nnu \\
\km1  \kp2    \et3  \et4 &\quad & - \facd{i}{4} A \ \nnu \\
\kp1  \bk2    \p03  \pm4 &\quad & \facd{i}{2\sqrt{2}} C \ \nnu \\
\kp1  \bk2    \et3  \pm4 &\quad & - \facd{i}{\sqrt{6}} A \ \nnu \\
\km1  \kp2    \et3  \p04 &\quad & - \facd{i}{2\sqrt{3}} A \ \nnu \\
\eea

\bea
\bk1 \, \kp2 \, \pm3 \, \pm4 \, \pp5 \, \p06 & \quad & \facq{i}{6\,\sqrt{2}} \left[ - \frac  {1}{2}
\Scat{1}{3}{4} +  \right. \ \nnu \\
& \quad & \left. \frac{1}{2} \Scat{2}{3}{4} + \sca{1}{6} - \sca{2}{6} \right] \ \nnu \\
\bk1 \, \kp2 \, \pm3 \, \p04 \, \p05 \, \p06 & \quad & \facq{i\,}{12\,\sqrt{2}} \left[ - \sca{1}{3} +
\sca{2}{3} +  \right. \ \nnu \\
& \quad &    \frac  {1}{3} \Scaq{1}{4}{5}{6} - \ \nnu \\
& \quad & \left. \frac  {1}{3} \Scaq{2}{4}{5}{6} \right] \ \nnu \\
\km1 \, \kp2 \, \pm3 \, \pm4 \, \pp5 \, \pp6 & \quad & \facq{i}{} \left[ -  \frac  {1}{90} \sca{1}{2} +
 \frac  {4}{90} \Scat{1}{3}{4} - \right. \ \nnu \\
& \quad &  \frac  {7}{180} \Scat{2}{3}{4} +  \frac  {2}{45} \sca{3}{4} - \ \nnu \\
& \quad &  \frac  {7}{180} \Scat{1}{5}{6} +  \frac  {4}{90} \Scat{2}{5}{6} - \ \nnu \\
& \quad &  \left.  \frac  {1}{40} \Scadd{3}{4}{5}{6} +  \frac  {2}{45} \sca{5}{6} \right] \ \nnu \\
\km1 \, \kp2 \, \pm3 \, \pp4 \, \p05 \, \p06 & \quad & \facq{i}{} \left[ - \frac  {1}{90} \sca{1}{2} +
 \frac  {2}{45} \sca{1}{3} - \right. \ \nnu \\
&\quad &  \frac  {7}{180} \sca{2}{3} -  \frac  {7}{180} \sca{1}{4} + \ \nnu \\
& \quad &  \frac  {2}{45} \sca{2}{4} -  \frac  {17}{180} \sca{3}{4} + \ \nnu \\
& \quad &  \frac  {1}{360} \Scat{1}{5}{6} + \frac{1}{360}\Scat{2}{5}{6} + \ \nnu \\
& \quad &  \frac  {4}{90} \Scat{3}{5}{6} +  \frac  {4}{90} \Scat{4}{5}{6} - \ \nnu \\
& \quad &  \left.  \frac  {17}{180} \sca{5}{6} \right] \ \nnu \\
\km1 \, \kp2 \, \p03 \, \p04 \, \p05 \, \p06 & \quad & \facq{i\,}{360} \left[ - \sca{1}{2} + \right. \ \nnu \\
& \quad &  \frac  {1}{4} \Scac{1}{3}{4}{5}{6} +  \ \nnu \\
& \quad &  \frac  {1}{4} \Scac{2}{3}{4}{5}{6} - \ \nnu \\
& \quad &   \frac  {1}{6} (\sca{3}{4} + \sca{3}{5} + \sca{3}{6} \ \nnu \\
& \quad &  \left. + \sca{4}{5} + \sca{4}{6} + \sca{5}{6} ) \right] \ \nnu \\
\k01 \, \km2 \, \pp3 \, \p04 \, \p05 \, \p06 & \quad & \facq{i\,}{12 \sqrt{2}} \left[
\sca{2}{3} - \sca{1}{3} - \right. \ \nnu \\
& \quad &  \frac  {1}{3} \Scaq{2}{4}{5}{6} +  \ \nnu \\
& \quad & \left. \frac  {1}{3} \Scaq{1}{4}{5}{6} \right] \ \nnu \\
\k01 \, \km2 \, \pm3 \, \pp4 \, \pp5 \, \p06 & \quad & \facq{i}{6 \, \sqrt{2}} \left[
 \frac  {1}{2} \Scat{2}{4}{5} - \right. \ \nnu \\
& \quad &  \frac  {1}{2} \Scat{1}{4}{5} -  \ \nnu \\
& \quad & \left. \sca{2}{6} + \sca{1}{6} \right] \ \nnu \\
\k01 \, \bk2 \, \pm3 \, \pm4 \, \pp5 \, \pp6 & \quad & \facq{i}{} \left[ -  \frac  {1}{90} \sca{1}{2} -
 \frac  {7}{180} \Scat{2}{3}{4} +  \right. \ \nnu \\
& \quad &  \frac  {4}{90} \Scat{1}{3}{4} +  \frac  {2}{45} \sca{3}{4} + \ \nnu \\
& \quad &  \frac  {4}{90} \Scat{2}{5}{6} -  \frac  {7}{180} \Scat{1}{5}{6} - \ \nnu \\
& \quad & \left.  \frac  {1}{40} \Scadd{3}{4}{5}{6} +  \frac  {2}{45} \sca{5}{6} \right] \ \nnu \\
\k01 \, \bk2 \, \pm3 \, \pp4 \, \p05 \, \p06 & \quad & \facq{i\,}{} \left[ - \frac  {1}{90} \sca{1}{2}
-  \frac  {7}{180} \sca{2}{3} + \right. \ \nnu \\
& \quad & \frac{2}{45} \sca{1}{3} + \frac  {2}{45} \sca{2}{4} - \frac  {7}{180} \sca{1}{4} - \ \nnu \\
& \quad &  \frac  {7}{180} \sca{1}{4} -  \frac  {17}{180} \sca{3}{4} + \ \nnu \\
& \quad &  \frac  {1}{360} \Scat{2}{5}{6} +  \frac  {1}{360} \Scat{1}{5}{6} + \ \nnu \\
& \quad &  \frac  {4}{90} \Scat{3}{5}{6} +  \frac  {4}{90} \Scat{4}{5}{6} - \ \nnu \\
& \quad & \left.  \frac  {17}{180} \sca{5}{6} \right] \ \nnu \\
\k01 \, \bk2 \, \p03 \, \p04 \, \p05 \, \p06 & \quad & \facq{i\,}{360} \left[
- \sca{1}{2}  + \right. \ \nnu \\
& \quad &  \frac  {1}{4} \Scac{2}{3}{4}{5}{6} +  \ \nnu \\
& \quad & \frac{1}{4} \Scac{1}{3}{4}{5}{6} - \ \nnu \\
& \quad &   \frac  {1}{6} ( \sca{3}{4} + \sca{3}{5} + \sca{3}{6} +  \ \nnu \\
& \quad & \left.   \sca{4}{5} +  \sca{4}{6} + \sca{5}{6})
\right] \ \nnu \\
\et1 \, \bk2 \, \kp3 \, \pm4 \, \pm5 \, \pp6 & \quad & \facq{i}{\sqrt{6}}
\left[ - \frac  {2}{15} \sca{2}{3} +   \frac  {1}{30} \sca{1}{2} + \right. \ \nnu \\
& \quad &  \frac  {1}{30} \sca{1}{3} +  \frac  {7}{60} \Scat{2}{4}{5} + \ \nnu \\
& \quad &  \frac  {7}{60} \Scat{3}{4}{5} -  \frac  {2}{15} \Scat{1}{4}{5} - \ \nnu \\
& \quad &  \frac  {2}{15} \sca{4}{5} -  \frac  {2}{15} \sca{2}{6}  - \ \nnu \\
& \quad &  \frac  {2}{15} \sca{3}{6} +  \frac  {1}{5} \sca{1}{6} + \ \nnu \\
& \quad &  \left.  \frac  {1}{30} \Scat{6}{4}{5} \right] \ \nnu \\
\et1 \, \bk2 \, \kp3 \, \pm4 \, \p05 \, \p06 & \quad & \facq{i}{\sqrt{6}} \left[
-  \frac  {1}{15} \sca{2}{3} +  \frac  {1}{60} \sca{1}{2} + \right. \ \nnu \\
& \quad &  \frac  {1}{60} \sca{1}{3} +  \frac  {11}{60} \sca{2}{4} + \ \nnu \\
& \quad &  \frac  {11}{60} \sca{3}{4} -  \frac  {7}{30} \sca{1}{4} - \ \nnu \\
& \quad &  \frac  {1}{15} \Scat{2}{5}{6} -  \frac  {1}{15} \Scat{3}{5}{6} + \ \nnu \\
& \quad &  \frac  {1}{10} \Scat{1}{5}{6} -  \frac  {1}{15} \Scat{4}{5}{6} + \ \nnu \\
& \quad & \left.  \frac  {1}{10} \sca{5}{6} \right] \ \nnu \\
\et1 \, \k02 \, \bk3 \, \pm4 \, \pp5 \, \p06 & \quad & \facq{i}{\sqrt{3}} \left[
 \frac  {1}{15} \sca{2}{3} -  \frac  {1}{60} \sca{1}{3} - \right.  \ \nnu \\
& \quad &  \frac  {1}{60} \sca{1}{2} +  \frac  {1}{15} \sca{3}{4} + \ \nnu \\
& \quad &  \frac  {1}{15} \sca{2}{4} -  \frac  {1}{10} \sca{1}{4} + \ \nnu \\
& \quad &  \frac  {1}{15} \sca{3}{5} +  \frac  {1}{15} \sca{2}{5} - \ \nnu \\
& \quad &  \frac  {1}{10} \sca{1}{5} -  \frac  {1}{10} \sca{4}{5} - \ \nnu \\
& \quad &  \frac  {11}{60} \sca{3}{6} -  \frac  {11}{60} \sca{2}{6} + \ \nnu \\
& \quad &  \frac  {7}{30} \sca{1}{6} +  \frac  {1}{15} \sca{4}{6} + \ \nnu \\
& \quad & \left.   \frac  {1}{15} \sca{5}{6} \right] \ \nnu \\
\et1 \, \k02 \, \bk3 \, \p04 \, \p05 \, \p06 & \quad & \facq{i}{\sqrt{3}} \left[
 \frac  {1}{30} \sca{2}{3} -  \frac  {1}{120} \sca{1}{3} - \right. \ \nnu \\
& \quad &  \frac  {1}{120} \sca{1}{2} -  \frac  {1}{120} \Scaq{3}{4}{5}{6} -  \ \nnu \\
& \quad &  \frac  {1}{120} \Scaq{2}{4}{5}{6}  +  \frac  {1}{180} \Scaq{1}{4}{5}{6} + \ \nnu \\
& \quad & \left.  \frac  {1}{180} ( \sca{5}{6} + \sca{4}{5} + \sca{4}{6} ) \right] \ \nnu \\
\eea
}

The relevant vertices coming from the mass term in the strong Lagrangian are:

{\footnotesize
\bea
\K0 \, \Bk \, \Kp \, \Km & \quad &  i\, \facd{m_K^2}{3} \ \nnu \\
\K0 \, \K0 \, \Bk \, \Bk & \quad & i\, \facd{m_K^2}{6} \ \nnu \\
\eta \, \eta \, \eta \, \eta & \quad & \facd{i}{27} ( 2 m_K^2 -  \frac  {7}{8} m_\pi^2 ) \ \nnu \\
\Kp \, \Kp \, \Km \, \Km  & \quad & i\, \facd{m_K^2}{6} \ \nnu \\
\Pm \, \Pm \,\Pp \,\Pp  &\quad & \facm{m^2_\pi}{6} \ \nnu \\
\Pp \,\Pm  \, \P0 \, \P0 &\quad & \facm{m^2_\pi}{6} \ \nnu \\
\P0 \, \P0 \,\P0 \,\P0 &\quad &  \facm{m^2_\pi}{24} \ \nnu \\
\eta \, \eta \,\Pp \, \Pm &\quad &  \facm{m^2_\pi}{6} \ \nnu \\
\eta \,\eta \,\P0 \,\P0 &\quad &  \facm{m^2_\pi}{12} \ \nnu \\
\Kp  \, \Km \,\Pp \,\Pm &\quad &  \facm{(m^2_\pi + m_K^2)}{6} \ \nnu \\
\Kp  \, \Km \,\P0 \,\P0 &\quad &  \facm{(m^2_\pi + m_K^2)}{12} \ \nnu \\
\eta \, \Bk \, \Kp \, \Pm & \quad & \facd{i\,}{ 3 \sqrt{6} } ( m_\pi^2 - m_K^2 ) \ \nnu \\
\eta \, \Kp \, \Km \, \P0 & \quad & \facd{i\,}{ 6\,\sqrt{3}} (m_\pi^2 - m_K^2 ) \ \nnu \\
\eta \, \K0 \, \Km \, \Pp & \quad & \facd{i\,}{ 3 \, \sqrt{6}} ( m_\pi^2 - m_K^2 ) \ \nnu \\
\eta \, \Bk \, \K0 \, \P0 & \quad & - \facd{i\,}{ 6\, \sqrt{3} } ( m_\pi^2 - m_K^2 ) \ \nnu \\
\K0 \,\Bk \, \Pp \,\Pm &\quad &  \facm{(m^2_\pi + m_K^2)}{6} \ \nnu \\
\K0 \,\Bk \, \P0 \,\P0 &\quad &  \facm{(m^2_\pi + m_K^2)}{12} \ \nnu \\
\eea

\bea
\Km \, \Kp \, \Pm \, \Pm \, \Pp \, \Pp & \quad & - \facq{i}{90} ( m_K^2 + 2 m_\pi^2 ) \ \nnu \\
\Km \, \Kp \, \Pm \, \Pp \, \P0 \, \P0 & \quad & - \facq{i}{90} ( m_K^2 + 2 m_\pi^2 ) \ \nnu \\
\Kp \, \Km \, \P0 \, \P0 \, \P0 \, \P0 & \quad & - \facq{i}{360} ( m_K^2 + 2 m_\pi^2 ) \ \nnu \\
\K0 \, \Bk \, \P0 \, \P0 \, \Pp \, \Pm & \quad & - \facq{i}{90}  ( m_K^2 + 2 m_\pi^2 ) \ \nnu \\
\K0 \, \Bk \, \Pp \, \Pm \, \Pp \, \Pm & \quad & - \facq{i}{90} ( m_K^2 + 2 m_\pi^2 ) \ \nnu \\
\K0 \, \Bk \, \P0 \, \P0 \, \P0 \, \P0 & \quad & - \facq{i}{360} (m_K^2 + 2 m_\pi^2) \ \nnu \\
\Bk \, \Kp \, \Pp \, \Pm \, \Pm \, \eta & \quad & - \facq{i}{15\,\sqrt{6}} m_\pi^2 \ \nnu \\
\Bk \, \Kp \, \P0 \, \P0 \, \Pm \, \eta & \quad & - \facq{i}{30\,\sqrt{6}} m_\pi^2 \ \nnu \\
\K0 \, \Km \, \Pp \, \Pp \, \Pm \, \eta & \quad &  - \facq{i}{30\,\sqrt{6}} m_\pi^2 \ \nnu \\
\K0 \, \Km \, \Pp \, \Pp \, \Pm \, \eta & \quad & - \facq{i}{15\,\sqrt{6}} m_\pi^2 \ \nnu \\
\Km \, \K0 \, \P0 \, \P0 \, \Pp \, \eta & \quad & - \facq{i}{30\,\sqrt{6}} m_\pi^2 \ \nnu \\
\K0 \, \Bk \, \P0 \, \Pp \, \Pm \, \eta & \quad & \facq{i}{30\,\sqrt{3}} m_\pi^2 \ \nnu \\
\K0 \, \Bk \, \P0 \, \P0 \, \P0 \, \eta & \quad & \facq{i}{60\,\sqrt{3}} m_\pi^2 \ \nnu \\
\eea
}

\section{Feynman Rules for the $\Delta S=1$ Chiral Lagrangian}

\vspace{.5cm}\noindent

The relevant vertices coming from the term $G_{LL}^a$ in the weak
chiral Lagrangian are:
{\footnotesize
 \bea
K^+ (p_1) \pi^- (p_2)
&\quad &    \frac  {2 i}{f^2} \ p_1 \cdot p_2 \  \nnu\\
\kp1 \, \kp2 \, \km3 \, \pm4  &\quad &
\facq{i}{} \left[ - \frac  {1}{3} \Scat{3}{1}{2} +  \frac  {4}{3} \sca{1}{2} + \right. \ \nnu \\
& \quad &  \left. \frac  {4}{3} \sca{3}{4} - \Scat{4}{1}{2} \right] \ \nnu \\
\kp1 \, \et2 \, \et3 \, \pm4 &\quad &
\facq{i}{} \left[  \frac  {1}{2} \Scat{4}{2}{3} - \sca{1}{4} \right] \ \nnu \\
\k01 \, \k02 \, \pp3 \, \km4 & \quad &
\facq{i}{} \left[ \sca{1}{2} -  \frac  {1}{2} \Scat{4}{1}{2} - \right. \ \nnu \\
& \quad & \left.  \frac  {1}{2} \Scat{3}{1}{2} + \sca{3}{4} \right] \ \nnu \\
\kp1 \, \pm2 \, \p03 \, \p04 & \quad &
\facq{i}{} \left[ -  \frac  {2}{3} \sca{1}{2} +  \frac  {1}{6} \Scat{1}{3}{4} - \right. \ \nnu \\
& \quad & \left. \frac  {1}{3} \Scat{2}{3}{4} + \sca{3}{4} \right] \ \nnu \\
\kp1 \, \pm2 \, \pm3 \, \pp4 & \quad &
\facq{i}{} \left[ - \Scat{1}{2}{3} +  \frac  {4}{3}  \sca{2}{3} +  \right. \ \nnu \\
& \quad & \left.   \frac  {4}{3} \sca{1}{4} -  \frac  {1}{3} \Scat{4}{2}{3} \right] \ \nnu
\\
\kp1 \, \k02 \, \bk3 \, \pm4 & \quad &
\facq{i}{} \left[  \frac  {4}{3} \sca{1}{3} -  \frac  {2}{3} \sca{1}{2} +
 \frac  {4}{3} \sca{3}{4} - \right. \ \nnu \\
& \quad & \left.  \frac{2}{3} \sca{2}{4} -  \frac{4}{3} \sca{1}{4} \right] \ \nnu \\
\k01 \, \km2 \, \kp3 \, \p04 & \quad &
\facq{i}{\sqrt{2}} \left[ - \sca{1}{3} + \sca{2}{3} - \sca{1}{4} + \sca{2}{4} \right] \ \nnu \\
\k01 \, \pm2 \, \pp3 \, \p04 & \quad &
 \facq{i\, \sqrt{2}}{} \left[ \sca{1}{2} - \sca{1}{4} - \sca{2}{4} + \sca{3}{4} \right] \ \nnu \\
\kp1 \, \pm2 \, \p03 \, \et4  & \quad &
\facq{i}{\sqrt{3}} \left[ - 2 \sca{2}{4} + \sca{1}{2} + 3 \sca{3}{4} - \right. \ \nnu \\
& \quad & \left. 3 \sca{1}{3} + \sca{2}{3} \right] \ \nnu \\
\k01 \, \pm2 \, \pp3 \, \et4 & \quad &
\sqrt{ \frac  {2}{3}} \facq{i}{} \left[ \sca{2}{4} - 2 \sca{1}{2} + \sca{2}{3} \right] \ \nnu \\
\k01 \, \kp2 \, \km3 \, \et4 & \quad &
\sqrt{ \frac  {3}{2}} \facq{i}{} \left[ - \sca{1}{4} + \sca{3}{4}  - 2 \sca{2}{4} + \right. \ \nnu \\
& \quad & \left. \frac  {5}{3} \sca{1}{2} -  \frac  {1}{3} \sca{2}{3} \right] \ \nnu \\
\eea

\bea
\kp1 \, \kp2 \, \km3 \, \pm4 \, \p05 \, \p06 &\quad&
\facs{i}{} \left[  \frac  {17}{360} \Scat{3}{1}{2} - \right. \ \nnu \\
& \quad & \frac  {43}{180} \sca{1}{2} -  \ \nnu \\
& \quad &  \frac  {43}{45} \sca{3}{4} +   \frac  {41}{120} \Scat{4}{1}{2}  +   \ \nnu \\
& \quad &   \frac  {37}{72} \Scat{3}{5}{6}  - \ \nnu \\
& \quad &   \frac  {11}{120} \Scadd{1}{2}{5}{6} + \ \nnu \\
& \quad &  \left.  \frac  {151}{360} \Scat{4}{5}{6} -  \frac  {13}{12} \sca{5}{6} \right] \ \nnu \\
\kp1 \, \kp2 \, \km3 \, \pm4 \, \pm5 \, \pp6 &\quad &
\facs{i}{} \left[  \frac  {31}{90} \Scat{3}{1}{2} -  \frac  {44}{45} \sca{1}{2} - \right. \ \nnu \\
 &\quad &    \frac  {28}{30} \Scat{3}{4}{5} +  \frac  {16}{20} \Scadd{1}{2}{4}{5}  \ \nnu \\
&\quad &   -  \frac  {44}{45} \sca{4}{5} +  \frac  {10}{9} \sca{3}{6} -  \ \nnu \\
 &\quad &  \left.   \frac  {28}{30} \Scat{6}{1}{2} +  \frac  {31}{90} \Scat{6}{4}{5} \right] \ \nnu \\
\kp1 \, \pm2 \, \p03 \, \p04 \, \p05 \, \p06 & \quad &
\facs{i}{} \left[  \frac  {4}{45} \sca{1}{2} - \right. \ \nnu \\
& \quad &  \frac  {13}{720} \Scac{1}{3}{4}{5}{6} +  \ \nnu \\
& \quad &     \frac  {8}{180} \Scac{2}{3}{4}{5}{6}  -  \ \nnu \\
& \quad &   \frac  {7}{192} ( \sca{3}{4} + \sca{3}{5} + \sca{3}{6} +   \ \nnu \\
& \quad &  \left. \left.  \sca{4}{5} +  \sca{4}{6} + \sca{5}{6} \right) \right] \ \nnu \\
\kp1 \, \pm2 \, \pm3 \, \pp4 \, \p05 \, \p06 &\quad&
\facs{i}{} \left[ \frac  {7}{20} \Scat{1}{2}{3} -  \frac  {149}{90} \sca{2}{3} - \right. \ \nnu \\
& \quad &  \frac  {13}{180} \Scat{1}{5}{6} + \ \nnu \\
& \quad &  \frac  {89}{120} \Scadd{2}{3}{5}{6} - \ \nnu \\
& \quad &  \frac  {11}{6} \sca{5}{6}  -  \frac  {22}{45} \sca{1}{4} +  \ \nnu \\
& \quad & \left.  \frac  {8}{90} \Scat{4}{2}{3} +  \frac  {5}{36} \Scat{4}{5}{6} \right] \ \nnu \\
\kp1 \, \pm2 \, \pm3 \, \pm4 \, \pp5 \, \pp6 & \quad & \facs{i}{}
\left[  \frac  {47}{135} \Scaq{1}{2}{3}{4} -
\right. \ \nnu \\
& \quad &  \frac  {64}{135} (\sca{2}{3} + \sca{2}{4} + \sca{3}{4} ) - \ \nnu \\
& \quad &  \frac  {44}{90} \Scat{1}{5}{6} +
\ \nnu \\
& \quad & \left.   \frac  {9}{30} \Scadt{5}{6}{2}{3}{4} -  \frac  {4}{9} \sca{5}{6} \right] \ \nnu \\
\k01 \, \k02 \, \bk3 \, \pm4 \, \pp5 \, \p06 & \quad & \facs{i}{\sqrt{2}} \left[  \frac  {1}{4}
\Scat{3}{1}{2} -  \frac  {1}{2} \sca{1}{2} + \right. \ \nnu \\
& \quad &  \frac  {3}{5} \sca{3}{4} -  \frac  {59}{60} \Scat{4}{1}{2} - \ \nnu \\
& \quad &  \frac  {7}{3} \sca{3}{6} +  \frac  {17}{12} \Scat{6}{1}{2} + \ \nnu \\
& \quad &  \frac  {19}{15} \sca{4}{6} +  \frac  {5}{6} \sca{3}{5} - \ \nnu \\
& \quad & \left.  \frac  {1}{12} \Scat{5}{1}{2} +  \frac  {1}{10} \sca{4}{5} -  \frac  {7}{6} \sca{5}{6}
\right] \ \nnu \\
\kp1 \, \et2 \, \et3 \, \pm4 \, \p05 \, \p06 & \quad & \facs{i}{} \left[ - \frac  {19}{120}
\Scat{4}{2}{3} +  \frac  {13}{60} \sca{1}{4} +  \right. \ \nnu \\
& \quad &  \frac  {1}{24} \Scadd{2}{3}{5}{6} + \ \nnu \\
& \quad & \frac  {1}{24} \Scat{1}{5}{6} + \ \nnu \\
& \quad & \left.  \frac  {1}{20} \Scat{4}{5}{6} -  \frac  {1}{4} \sca{5}{6} \right] \ \nnu \\
\kp1 \, \et2 \, \et3 \, \pm4 \, \pm5 \, \pp6 & \quad & \facs{i}{} \left[
-  \frac  {29}{120} \Scadd{2}{3}{4}{5} + \right. \ \nnu \\
& \quad &  \frac  {19}{30} \Scat{1}{4}{5} - \ \nnu \\
& \quad &  \frac  {2}{5} \sca{4}{5} +  \frac  {1}{3} \Scat{6}{2}{3} - \ \nnu \\
& \quad & \left.  \frac  {2}{3} \sca{1}{6} +  \frac  {1}{20} \Scat{6}{4}{5} \right] \ \nnu \\
\k01 \, \pm2 \, \pp3 \, \p04 \, \p05 \, \p06 & \quad & \facs{i}{\sqrt{2}} \left[ -  \frac  {5}{6}
\sca{1}{2} + \right. \ \nnu \\
& \quad & \frac  {5}{18} \Scaq{1}{4}{5}{6} +  \ \nnu \\
& \quad &  \frac  {5}{18} \Scaq{2}{4}{5}{6} - \ \nnu \\
& \quad &  \frac  {1}{9} ( \sca{4}{5} + \sca{4}{6} + \sca{5}{6} ) -
 \ \nnu \\
& \quad &  \left.  \frac  {1}{6} \Scaq{3}{4}{5}{6} \right] \ \nnu \\
\k01 \, \pm2 \, \pm3 \, \pp4 \, \pp5 \, \p06 & \quad & \facs{i}{\sqrt{2}} \left[ -  \frac  {5}{6}
\Scat{1}{2}{3} - \sca{2}{3} + \right. \ \nnu \\
& \quad &  \frac  {5}{3} \sca{1}{6} +  \frac  {5}{3} \Scat{6}{2}{3} + \ \nnu \\
& \quad &  \frac  {1}{12} \Scadd{2}{3}{4}{5} -  \ \nnu \\
& \quad & \left. 2 \Scat{6}{4}{5} + \frac  {4}{3} \sca{4}{5} \right] \ \nnu \\
\k01 \, \km2 \, \kp3 \, \p04 \, \p05 \, \p06 & \quad & \facs{i}{\sqrt{2}} \left[
 \frac  {1}{6} \sca{1}{3} -  \frac  {1}{6} \sca{2}{3} + \right. \ \nnu \\
& \quad &  \frac  {1}{9} \Scaq{1}{4}{5}{6} - \ \nnu \\
& \quad & \left.  \frac{1}{9} \Scaq{2}{4}{5}{6} \right] \ \nnu \\
\k01 \, \km2 \, \kp3 \, \pm4 \, \pp5 \, \p06 & \quad & \facs{i}{\sqrt{2}} \left[
-  \frac  {1}{2} \sca{1}{2} + \right. \ \nnu \\
& \quad &  \frac  {4}{3} \sca{1}{3} -  \frac  {5}{6} \sca{2}{3} -  \ \nnu \\
& \quad &  \frac  {21}{5} \sca{1}{4} +  \frac  {79}{30} \sca{2}{4} -  \frac  {1}{5} \sca{3}{4} + \ \nnu \\
& \quad & 4 \sca{1}{6} -  \frac  {7}{6} \sca{2}{6} -  \frac  {4}{3} \sca{3}{6} + \ \nnu \\
& \quad &  \frac  {22}{15} \sca{4}{6} +  \frac  {1}{6} \sca{1}{5} - \frac{1}{3} \sca{2}{5} +
 \frac  {5}{6} \sca{3}{5} + \ \nnu \\
& \quad & \left.  \frac  {3}{10} \sca{4}{5} -  \frac  {13}{6} \sca{5}{6} \right] \ \nnu \\
\kp1 \, \k02 \, \bk3 \, \pm4 \, \p05 \, \p06 & \quad & \facs{i}{} \left[
-  \frac  {43}{180} \sca{1}{3} +  \frac{17}{180} \sca{1}{2} -  \right.
\ \nnu \\
& \quad & \frac{16}{45} \sca{3}{4} +  \frac{43}{90} \sca{2}{4} + \ \nnu \\
& \quad &   \frac  {73}{180} \sca{1}{4} -  \frac  {5}{72} \Scat{3}{5}{6} -
\ \nnu \\
& \quad &  \frac  {11}{72} \Scat{2}{5}{6} +  \frac  {19}{360} \Scat{1}{5}{6} + \ \nnu \\
& \quad &  \left.  \frac  {7}{360} \Scat{4}{5}{6} -  \frac  {1}{12} \sca{5}{6} \right] \ \nnu \\
\kp1 \, \k02 \, \bk3 \, \pm4 \, \pm5 \, \pp6 & \quad &
\facs{i}{} \left[ -  \frac  {1}{2} \sca{2}{3} - \right. \ \nnu \\
& \quad &  \frac  {44}{45} \sca{1}{3} +  \frac  {107}{90} \sca{1}{2} -  \ \nnu \\
& \quad &  \frac  {1}{15} \Scat{3}{4}{5} -  \frac  {1}{15} \Scat{2}{4}{5} + \ \nnu \\
& \quad &  \frac  {9}{10} \Scat{1}{4}{5} - \ \nnu \\
& \quad &  \frac  {26}{45} \sca{4}{5} +  \frac  {17}{18} \sca{3}{6} -
\ \nnu \\
& \quad &  \frac  {2}{9} \sca{2}{6} -  \frac  {133}{90} \sca{1}{6} + \ \nnu \\
& \quad & \left. \frac  {4}{90} \Scat{6}{4}{5} \right] \ \nnu \\
\k01 \, \et2 \, \et3 \, \pm4 \, \pp5 \, \p06 & \quad & \facs{i}{\sqrt{2}} \left[  \frac  {1}{2}
\Scat{4}{2}{3} -  \frac  {3}{2} \sca{1}{4} - \right. \ \nnu \\
& \quad &  \frac  {1}{2} \Scat{6}{2}{3} +  \frac  {3}{2} \sca{1}{6} + \ \nnu \\
& \quad & \left.   \frac  {1}{2} \sca{4}{6} -  \frac  {1}{2} \sca{5}{6}  \right] \ \nnu \\
\eea
}

 \vspace{.5cm}\noindent
The relevant vertices coming from the term $G_{LL}^b$ in the weak
chiral Lagrangian are:
{\footnotesize
\bea
K^0 (p_1) \pi^0 (p_2)
&\quad &    \frac  {i \sqrt{2}}{f^2} \ p_1  \cdot  p_2 \  \nnu\\
K^0 (p_1) \eta (p_2)
&\quad &    \frac  {i}{f^2} \sqrt{ \frac  {2}{3}} \ p_1  \cdot  p_2 \  \nnu\\
\kp1 \, \kp2 \, \km3 \, \pm4 & \quad & \facq{i}{} \left[ -  \frac  {1}{2} \Scat{3}{1}{2} +
\sca{1}{2} + \right. \ \nnu \\
& \quad & \left. \sca{3}{4} -  \frac  {1}{2} \Scat{4}{1}{2} \right] \ \nnu \\
\kp1 \, \pm2  \, \et3 \, \et4 & \quad & \facq{i}{} \left[  \frac  {1}{3} \sca{3}{4} -
\Scat{1}{3}{4} + \right. \ \nnu \\
& \quad & \left.  \frac  {1}{6} \Scat{2}{3}{4} \right]  \ \nnu \\
\k01 \, \p02 \, \et3 \, \et4 & \quad & \facq{i}{\sqrt{2}} \left[ - \frac  {1}{3} \sca{3}{4} +
 \frac  {2}{3} \Scat{1}{3}{4} + \right. \ \nnu \\
& \quad & \left.  \frac  {1}{3} \Scat{2}{3}{4} - \sca{1}{2} \right] \ \nnu \\
\k01 \, \k02 \, \bk3 \, \p04 & \quad & \facq{i\,\sqrt{2}}{3} \left[ 2 \sca{3}{4} -
 \frac  {1}{3} \Scat{4}{1}{2} \right] \ \nnu \\
\k01 \, \k02 \, \km3 \, \pp4 & \quad & \facq{i}{} \left[  \frac  {4}{3} \sca{1}{2} -
 \frac  {1}{3} \Scat{3}{1}{2} - \right. \ \nnu \\
& \quad & \left.  \frac  {1}{3} \Scat{4}{1}{2} \right] \ \nnu \\
\kp1 \, \pm2 \, \p03 \, \p04 & \quad & \facq{i}{} \left[ -  \frac  {1}{2} \Scat{2}{3}{4} +
\sca{3}{4} \right] \ \nnu \\
\kp1 \, \pm2 \, \pm3 \, \pp4 & \quad & \facq{i}{} \left[ -  \frac  {1}{2} \Scat{1}{2}{3} +
\sca{2}{3} + \right. \ \nnu \\
& \quad & \left. \sca{1}{4} -  \frac  {1}{2} \Scat{4}{2}{3} \right] \ \nnu \\
\k01 \, \bk2 \, \kp3 \, \pm4 & \quad & \facq{i}{} \left[  \frac  {4}{3} \sca{1}{2} -
 \frac  {2}{3} \sca{1}{3} -  \frac  {2}{3} \sca{1}{4} \right] \ \nnu \\
\k01 \, \kp2 \, \km3 \, \p04 & \quad & \facq{i}{\sqrt{2}} \left[
 \frac  {5}{3} \sca{1}{3} -  \frac  {1}{3} \sca{1}{2} - \right. \ \nnu \\
& \quad & 2 \sca{1}{4} +  \frac  {1}{3} \sca{3}{4} + \ \nnu \\
& \quad & \left.  \frac  {1}{3} \sca{2}{4}  \right] \ \nnu \\
\k01 \, \p02 \, \p03 \, \p04 & \quad & \facq{i}{\sqrt{2}}  \left[
-  \frac  {1}{9} \Scaq{1}{2}{3}{4} + \right. \ \nnu \\
& \quad & \left.  \frac  {1}{9} ( \sca{2}{3} + \sca{3}{4} + \sca{2}{4} ) \right] \ \nnu \\
\k01 \, \pp2 \, \pm3 \, \p04 & \quad & \facq{i}{\sqrt{2}} \left[
 \frac  {7}{3} \sca{1}{3} -  \frac  {10}{3} \sca{1}{4} - \right. \ \nnu \\
& \quad & \left.  \frac  {5}{3} \sca{3}{4} +  \frac  {1}{3} \sca{1}{2} +  \frac  {7}{3} \sca{2}{4}
\right] \ \nnu \\
\kp1 \, \pm2 \, \p03 \, \et4 & \quad & \facq{i}{\sqrt{3}} \left[ - \sca{2}{4} +
2 \sca{3}{4} - \right. \ \nnu \\
& \quad & \left. 2 \sca{1}{3} + \sca{2}{3} \right] \ \nnu \\
\k01 \, \p02 \, \p03 \, \et4 & \quad & \facq{i}{\sqrt{6}} \left[ -  \frac  {1}{3}
\sca{1}{4} -  \frac  {1}{3} \Scat{4}{2}{3} +  \right. \ \nnu \\
& \quad & \left. \Scat{1}{2}{3} - \sca{2}{3} \right] \ \nnu \\
\k01 \, \pm2 \, \pp3 \, \et4 & \quad & \facq{i}{\sqrt{6}} \left[ -  \frac  {2}{3} \sca{1}{4} +
 \frac  {7}{3} \sca{2}{4} - \right. \nnu \\
& \quad & \left. 3 \sca{1}{2} -  \frac  {5}{3} \sca{3}{4} + 3 \sca{1}{3} \right] \ \nnu \\
\eea

\bea
\kp1 \, \kp2 \, \km3 \, \pm4 \, \p05 \, \p06 & \quad & \facs{i}{} \left[  \frac  {1}{12}
\Scat{3}{1}{2} -  \frac  {1}{6} \sca{1}{2} - \right. \ \nnu \\
& \quad & \sca{3}{4} +  \frac  {1}{6} \Scat{4}{1}{2} + \ \nnu \\
& \quad &  \frac  {31}{60} \Scat{3}{5}{6} -  \ \nnu \\
& \quad &  \frac  {4}{60} \Scadd{1}{2}{5}{6} + \ \nnu \\
& \quad & \left.  \frac  {13}{30} \Scat{4}{5}{6} -  \frac  {29}{30} \sca{5}{6} \right] \ \nnu \\
\kp1 \, \kp2 \, \km3 \, \pm4 \, \pm5 \, \pp6 & \quad & \facs{i}{} \left[
 \frac  {5}{12} \Scat{3}{1}{2} -  \frac  {5}{6} \sca{1}{2} - \right. \ \nnu \\
& \quad &  \frac  {11}{12} \Scat{3}{4}{5} + \ \nnu \\
& \quad &  \frac  {8}{12} \Scadd{1}{2}{4}{5} - \ \nnu \\
& \quad &  \frac  {5}{6} \sca{4}{5} + \sca{3}{6} - \ \nnu \\
& \quad & \left.  \frac  {11}{12} \Scat{6}{1}{2} +  \frac  {5}{12} \Scat{6}{4}{5} \right] \ \nnu \\
\kp1 \, \p02 \, \p03 \, \p04 \, \p05 \, \pm6 & \quad & \facs{i}{} \left[
 \frac  {1}{16} \Scac{6}{2}{3}{4}{5} - \right. \ \nnu \\
& \quad &  \frac  {1}{24} ( \sca{2}{3} + \sca{2}{4} + \sca{2}{5} + \ \nnu \\
& \quad & \left. \sca{3}{4} + \sca{3}{5} + \sca{4}{5} ) \right] \ \nnu \\
\kp1 \, \pm2 \, \pm3 \, \pp4 \, \p05 \, \p06 & \quad & \facs{i}{} \left[
 \frac  {5}{24} \Scat{1}{2}{3} -  \frac  {19}{12} \sca{2}{3} + \right. \ \nnu \\
& \quad &  \frac  {3}{4} \Scadd{2}{3}{5}{6} - \ \nnu \\
& \quad &  \frac  {11}{6} \sca{5}{6} -  \frac  {5}{12} \sca{1}{4} + \ \nnu \\
& \quad & \left.  \frac  {1}{8} \Scat{4}{2}{3} +  \frac  {1}{12} \Scat{4}{5}{6}  \right] \ \nnu \\
\kp1 \, \pm2 \, \pm3 \, \pm4 \, \pp5 \, \pp6 & \quad & \facs{i}{} \left[
 \frac  {5}{18} \Scaq{1}{2}{3}{4} - \right. \ \nnu \\
& \quad &  \frac  {4}{9} ( \sca{2}{3} + \sca{2}{4} + \sca{3}{4} ) - \ \nnu \\
& \quad &  \frac  {5}{12} \Scat{1}{5}{6} + \ \nnu \\
& \quad &  \frac  {11}{36} \Scadt{5}{6}{2}{3}{4} - \ \nnu \\
& \quad & \left. -  \frac  {1}{2} \sca{5}{6} \right] \ \nnu \\
\kp1 \, \pm2 \, \p03 \, \p04 \, \et5 \, \et6 & \quad & \facs{i}{} \left[
-  \frac  {1}{60} \sca{5}{6} +  \frac  {1}{30} \Scat{1}{5}{6} - \right. \ \nnu \\
& \quad &  \frac  {11}{120} \Scat{2}{5}{6} + \ \nnu \\
& \quad &  \frac  {2}{60} \Scadd{3}{4}{5}{6} + \ \nnu \\
& \quad & \left.  \frac  {1}{8} \Scat{2}{3}{4} -  \frac  {1}{4} \sca{3}{4} \right] \ \nnu \\
\kp1 \, \pm2 \, \pm3 \, \pp4 \, \et5 \, \et6 & \quad & \facs{i}{}
\left[ -  \frac  {1}{30} \sca{5}{6} +  \frac  {2}{30} \Scat{1}{5}{6} - \right. \ \nnu \\
& \quad &  \frac  {11}{60} \Scadd{5}{6}{2}{3} + \ \nnu \\
& \quad &  \frac  {3}{8} \Scat{1}{2}{3} -  \frac  {1}{4} \sca{2}{3} + \ \nnu \\
& \quad &  \frac  {19}{60} \Scat{4}{5}{6} -  \frac  {3}{4} \sca{1}{4} + \ \nnu \\
& \quad & \left.  \frac  {1}{8} \Scat{4}{2}{3} \right] \ \nnu \\
\k01 \, \p02 \, \p03 \, \p04 \, \et5 \, \et6 & \quad & \facs{i}{\sqrt{2}} \left[
 \frac  {1}{60} \sca{5}{6} -  \frac  {1}{30} \Scat{1}{5}{6} - \right. \ \nnu \\
& \quad &  \frac  {1}{60} \Scadt{5}{6}{2}{3}{4} + \ \nnu \\
& \quad &  \frac  {3}{30} \Scaq{1}{2}{3}{4} - \ \nnu \\
& \quad & \left.  \frac  {3}{60} ( \sca{2}{3} + \sca{2}{4} + \sca{3}{4} ) \right] \ \nnu \\
\k01 \, \pm2 \, \pp3 \, \p04 \, \et5 \, \et6 & \quad & \facs{i}{\sqrt{2}} \left[
 \frac  {1}{30} \sca{5}{6} -  \frac  {1}{15} \Scat{1}{5}{6}  + \right. \ \nnu \\
& \quad &  \frac  {31}{60} \Scat{2}{5}{6} -  \frac  {17}{12} \sca{1}{2} - \ \nnu \\
& \quad &  \frac  {19}{30} \Scat{4}{5}{6}  +  \frac  {29}{15} \sca{1}{4} + \ \nnu \\
& \quad &  \frac  {7}{20} \sca{2}{4} +  \frac  {1}{60} \Scat{3}{5}{6} + \ \nnu \\
& \quad & \left.  \frac  {1}{12} \sca{1}{3} -  \frac  {13}{20} \sca{3}{4} \right] \ \nnu \\
\k01 \, \k02 \, \bk3 \, \p04 \, \p05 \, \p06 & \quad & \facs{i}{\sqrt{2}} \left[
-  \frac  {1}{45} \Scaq{3}{4}{5}{6} + \right. \ \nnu \\
& \quad &  \frac  {1}{30} \Scadt{1}{2}{4}{5}{6} - \ \nnu \\
& \quad & \left.  \frac  {2}{45} ( \sca{4}{5} + \sca{4}{6} + \sca{5}{6} ) \right] \ \nnu \\
\k01 \, \k02 \, \bk3 \, \pm4 \, \pp5 \, \p06 & \quad & \facs{i}{\sqrt{2}} \left[
 \frac  {1}{4} \Scat{3}{1}{2} -  \frac  {1}{2} \sca{1}{2} + \right. \ \nnu \\
& \quad &  \frac  {8}{9} \sca{3}{4} -  \frac  {37}{36} \Scat{4}{1}{2} - \ \nnu \\
& \quad &  \frac  {253}{90} \sca{3}{6} +  \frac  {319}{180} \Scat{6}{1}{2} + \ \nnu \\
& \quad &  \frac  {13}{10} \sca{4}{6} +  \frac  {8}{9} \sca{3}{5} - \ \nnu \\
& \quad & \left.  \frac  {7}{36} \Scat{5}{1}{2} -  \frac  {41}{30} \sca{5}{6} \right] \ \nnu \\
\k01 \, \p02 \, \p03 \, \p04 \, \p05 \, \p06 & \quad & \facs{i}{60 \, \sqrt{2}} \left[
 \frac  {1}{5} \Scas{1}{2}{3}{4}{5}{5}{6} - \right. \  \nnu \\
& \quad &  \frac  {1}{10} ( \sca{2}{3} + \sca{2}{4} + \sca{2}{5} + \ \nnu \\
& \quad & \sca{2}{6} + \sca{3}{4} + \sca{3}{5} + \ \nnu \\
& \quad & \sca{3}{6} + \sca{4}{5} + \sca{4}{6} + \ \nnu \\
& \quad & \left. \sca{5}{6} ) \right] \ \nnu \\
\k01 \, \pm2 \, \pp3 \, \p04 \, \p05 \, \p06 & \quad & \facs{i}{\sqrt{2}} \left[
 - \frac  {163}{180} \sca{1}{2} +  \frac  {47}{135} \Scaq{1}{4}{5}{6} + \right. \ \nnu \\
& \quad &  \frac  {157}{540} \Scaq{2}{4}{5}{6} - \ \nnu \\
& \quad &  \frac  {43}{270} ( \sca{4}{5} + \sca{4}{6} + \sca{5}{6} ) - \ \nnu \\
& \quad & \left.  \frac  {13}{180} \sca{1}{3} -  \frac  {83}{540} \Scaq{3}{4}{5}{6} \right] \ \nnu \\
\k01 \, \pm2 \, \pm3 \, \pp4 \, \pp5 \, \p06 & \quad & \facs{i}{\sqrt{2}} \left[
-  \frac  {163}{180} \Scat{1}{2}{3} -  \frac  {17}{18} \sca{2}{3} + \right. \ \nnu \\
& \quad &  \frac  {91}{45} \sca{1}{6} +  \frac  {287}{180} \Scat{6}{2}{3} - \ \nnu \\
& \quad &  \frac  {13}{180} \Scat{1}{4}{5} + \ \nnu \\
& \quad &  \frac  {1}{9} \Scadd{2}{3}{4}{5} - \ \nnu \\
& \quad & \left.  \frac  {373}{180} \Scat{6}{4}{5} +  \frac  {25}{18} \sca{4}{5} \right] \ \nnu \\
\k01 \, \kp2 \, \km3 \, \p04 \, \p05 \, \p06 & \quad & \facs{i}{\sqrt{2}} \left[
-  \frac  {14}{45} \sca{1}{3} +  \frac  {1}{45} \sca{1}{2} + \right. \ \nnu \\
& \quad &  \frac  {59}{270} \Scaq{1}{4}{5}{6} - \ \nnu \\
& \quad &  \frac  {17}{270} \Scaq{3}{4}{5}{6} - \ \nnu \\
& \quad &  \frac  {1}{135} \Scaq{2}{4}{5}{6} - \ \nnu \\
& \quad & \left.  \frac  {7}{135} ( \sca{4}{5} + \sca{4}{6} + \sca{5}{6} ) \right] \ \nnu \\
\k01 \, \kp2 \, \km3 \, \pm4 \, \pp5 \, \p06 & \quad & \facs{i}{\sqrt{2}} \left[
-  \frac  {101}{90} \sca{1}{3} +  \frac  {47}{45} \sca{1}{2} - \right. \ \nnu \\
& \quad &  \frac  {1}{2} \sca{2}{3} -  \frac  {391}{90} \sca{1}{4} + \ \nnu \\
& \quad &   \frac  {7}{3} \sca{3}{4} -  \frac  {1}{6} \sca{2}{4} + \ \nnu \\
& \quad &  \frac  {24}{5} \sca{1}{6} -  \frac  {67}{90} \sca{3}{6} - \ \nnu \\
& \quad &  \frac  {71}{45} \sca{2}{6} +  \frac  {103}{90} \sca{4}{6} + \ \nnu \\
& \quad &  \frac  {7}{45} \sca{1}{5} -  \frac  {1}{6} \sca{3}{5} + \ \nnu \\
& \quad & \sca{2}{5} +  \frac  {1}{6} \sca{4}{5} - \ \nnu \\
& \quad & \left.  \frac  {91}{45} \sca{5}{6} \right] \ \nnu \\
\k01 \, \bk2 \, \kp3 \, \pm4 \, \p05 \, \p06 & \quad & \facs{i}{} \left[
-  \frac  {43}{180} \sca{1}{2} +  \frac  {17}{180} \sca{1}{3} + \right. \ \nnu \\
& \quad &  \frac  {107}{180} \sca{1}{4} -  \frac  {77}{360} \Scat{2}{5}{6} - \ \nnu \\
& \quad &  \frac  {7}{40} \Scat{1}{5}{6} +  \frac  {43}{360} \Scat{3}{5}{6} + \ \nnu \\
& \quad & \left.  \frac  {43}{360} \Scat{4}{5}{6} -  \frac  {3}{20} \sca{5}{6} \right] \
\nnu \\
\k01 \, \bk2 \, \kp3 \, \pm4 \, \pm5 \, \pp6 & \quad & \facs{i}{} \left[
-  \frac  {44}{45} \sca{1}{2} -  \frac  {1}{2} \sca{2}{3} + \right. \ \nnu \\
& \quad &  \frac  {107}{90} \sca{1}{3} +  \frac  {1}{36} \Scat{2}{4}{5} - \ \nnu \\
& \quad &  \frac  {1}{180} \Scat{1}{4}{5} +  \frac  {19}{36} \Scat{3}{4}{5} - \ \nnu \\
& \quad &  \frac  {11}{18} \sca{4}{5} +  \frac  {8}{9} \sca{2}{6} - \ \nnu \\
& \quad &  \frac  {1}{5} \sca{1}{6} -  \frac  {23}{18} \sca{3}{6} + \ \nnu \\
& \quad & \left.  \frac  {7}{36} \Scat{6}{4}{5} \right] \ \nnu \\
\eea } \vspace{.5cm} \noindent The relevant vertices coming from
the term $G_{\underline{8}}$ in the weak chiral Lagrangian are:
{\footnotesize
\bea
K^0 (p_1) \pi^0 (p_2)
&\quad &    \frac  {i \sqrt{2}}{f^2} \ p_1  \cdot  p_2 \  \nnu\\
K^0 (p_1) \eta (p_2)
&\quad &    \frac  {i}{f^2} \sqrt{ \frac  {2}{3}} \ p_1  \cdot  p_2 \  \nnu\\
K^+ (p_1) \pi^- (p_2)
&\quad &  -  \frac  {2 i}{f^2} \ p_1  \cdot  p_2 \  \nnu\\
\kp1 \, \p02 \, \p03 \, \pp4 & \quad & \facq{i}{} \left[  \frac  {2}{3} \sca{1}{4} -
 \frac  {1}{6} \Scat{1}{2}{3} -  \frac  {1}{6} \Scat{4}{2}{3} \right] \ \nnu \\
\kp1 \, \pm2 \, \pm3 \, \pp4 & \quad & \facq{i}{} \left[
 \frac  {1}{2} \Scat{1}{2}{3} -  \frac  {1}{3} \sca{2}{3} - \right. \ \nnu \\
& \quad & \left.  \frac  {1}{3} \sca{1}{4} -  \frac  {1}{6} \Scat{4}{2}{3} \right] \ \nnu \\
\k01 \, \p02 \, \p03 \, \p04 & \quad & \facq{i}{\sqrt{2}} \left[
-  \frac  {1}{9} \Scaq{1}{2}{3}{4} + \right. \ \nnu \\
& \quad & \left.  \frac  {1}{9} ( \sca{2}{3} + \sca{2}{4} + \sca{3}{4} )
\right] \ \nnu \\
\k01 \, \p02 \, \pp3 \, \pm4 & \quad & \facq{i}{\sqrt{2}} \left[  \frac  {1}{3} \sca{1}{4} -  \frac  {4}{3}
\sca{1}{2} +
\right. \ \nnu \\
& \quad &  \left.  \frac  {1}{3} \sca{2}{4} +  \frac  {1}{3} \sca{1}{3} +  \frac  {1}{3} \sca{2}{3} \right] \ \nnu \\
\kp1 \, \kp2 \, \km3 \, \pm4 & \quad & \facq{i}{} \left[ -  \frac  {1}{6} \Scat{3}{1}{2} -  \frac  {1}{3} \sca{1}{2}
-  \frac  {1}{3} \sca{3}{4} + \right. \ \nnu \\
& \quad & \left.   \frac  {1}{2} \Scat{4}{1}{2} \right] \ \nnu \\
\kp1 \, \pm2 \, \et3 \, \et4 & \quad & \facq{i}{} \left[  \frac  {1}{3} \sca{3}{4} -  \frac  {1}{3} \Scat{1}{3}{4} -
\Scat{2}{3}{4} + \sca{1}{2} \right] \ \nnu \\
\k01 \, \p02 \, \et3 \, \et4 & \quad & \facq{i}{\sqrt{2}} \left[ -  \frac  {1}{3} \sca{3}{4} +  \frac  {1}{3}
\Scat{1}{3}{4} + \right. \ \nnu \\
& \quad & \left.  \frac  {1}{3} \Scat{2}{3}{4} - \sca{1}{2} \right] \ \nnu \\
\k01 \, \k02 \, \bk3 \, \p04 & \quad & \facq{i \sqrt{2}}{3} \left[ 2 \sca{3}{4} - \Scat{4}{1}{2} \right] \ \nnu \\
\k01 \, \k02 \, \km3 \, \pp4 & \quad & \facq{i}{} \left[  \frac  {1}{3} \sca{1}{2} +
 \frac  {1}{6} \Scat{3}{1}{2} +  \frac  {1}{6} \Scat{4}{1}{2} - \sca{3}{4} \right] \ \nnu \\
\kp1 \, \k02 \, \bk3 \, \pm4 & \quad & \facq{i}{} \left[  \frac  {4}{3} \sca{2}{3} -  \frac  {4}{3} \sca{1}{3}
- \frac{4}{3} \sca{3}{4}
+  \frac  {4}{3} \sca{1}{4} \right] \ \nnu \\
\k01 \, \kp2 \, \km3 \, \p04 & \quad & \facq{i}{\sqrt{2}} \left[  \frac  {5}{3} \sca{1}{3} +
 \frac  {2}{3} \sca{1}{2} - \right. \ \nnu \\
& \quad &  \sca{2}{3} - \sca{1}{4} - \ \nnu \\
& \quad & \left.  \frac  {2}{3} \sca{3}{4} +  \frac  {1}{3} \sca{2}{4} \right] \ \nnu \\
\kp1 \, \pm2 \, \p03 \, \et4 & \quad & \facq{i}{\sqrt{3}} \left[ \sca{2}{4} - \sca{1}{2} -
\sca{3}{4} + \sca{1}{3} \right] \ \nnu \\
\k01 \, \p02 \, \p03 \, \et4 & \quad & \facq{i}{\sqrt{6}} \left[
-  \frac  {1}{3} \sca{1}{4} -  \frac  {1}{3} \Scat{4}{2}{3} + \right. \ \nnu \\
& \quad & \left. \Scat{1}{2}{3} - \sca{2}{3} \right] \ \nnu \\
\k01 \, \pp2 \, \pm3 \, \et4 & \quad & \facq{i}{\sqrt{6}} \left[ -  \frac  {2}{3} \sca{1}{4} +
 \frac  {1}{3}  \sca{3}{4} + \right. \ \nnu \\
& \quad & \sca{1}{3} -  \frac  {5}{3} \sca{2}{4} + \ \nnu \\
& \quad & \left. 3 \sca{1}{2} - 2 \sca{2}{3} \right] \ \nnu \\
\eea

\bea
\kp1 \, \kp2 \, \km3 \, \pm4 \, \p05 \, \p06 & \quad & \facs{i}{} \left[  \frac  {13}{360} \Scat{3}{1}{2} +
 \frac  {13}{180} \sca{1}{2} - \right. \ \nnu \\
& \quad &   \frac  {2}{45} \sca{3}{4} -  \frac  {7}{40} \Scat{4}{1}{2} + \ \nnu \\
& \quad &  \frac  {1}{360} \Scat{3}{5}{6} + \ \nnu \\
& \quad &  \frac  {1}{40} \Scadd{1}{2}{5}{6} + \ \nnu \\
& \quad & \left.  \frac  {1}{72} \Scat{4}{5}{6} +  \frac  {7}{60} \sca{5}{6} \right] \ \nnu \\
\kp1 \, \kp2 \, \km3 \, \pm4 \, \pm5 \, \pp6 & \quad & \facs{i}{} \left[
 \frac  {13}{180} \Scat{3}{1}{2} +  \frac  {13}{90} \sca{1}{2} + \right. \ \nnu \\
& \quad &  \frac  {1}{60} \Scat{3}{4}{5} - \ \nnu \\
& \quad &  \frac  {8}{60} \Scadd{1}{2}{4}{5} + \ \nnu \\
& \quad &  \frac  {13}{90} \sca{4}{5} -  \frac  {1}{9} \sca{3}{6} + \ \nnu \\
& \quad & \left.  \frac  {1}{60} \Scat{6}{1}{2} +  \frac  {13}{180} \Scat{6}{4}{5} \right] \ \nnu \\
\kp1 \, \pm2 \, \p03 \, \p04 \, \p05 \, \p06 & \quad & \facs{i}{} \left[ -  \frac  {4}{45} \sca{1}{2} + \right. \
\nnu \\
& \quad &  \frac  {13}{720} \Scac{1}{3}{4}{5}{6} + \ \nnu \\
& \quad &  \frac  {13}{720} \Scac{2}{3}{4}{5}{6} - \ \nnu \\
& \quad &  \frac  {1}{108} ( \sca{3}{4} + \sca{3}{5} + \sca{3}{6} + \ \nnu \\
& \quad & \left. \sca{4}{5} + \sca{4}{6} + \sca{5}{6} ) \right] \ \nnu \\
\kp1 \, \pm2 \, \pm3 \, \pp4 \, \p05 \, \p06 & \quad & \facs{i}{} \left[
-  \frac  {17}{120} \Scat{1}{2}{3} +  \frac  {13}{180} \sca{2}{3} + \right. \ \nnu \\
& \quad &  \frac  {13}{180} \Scat{1}{5}{6}  + \ \nnu \\
& \quad &  \frac  {1}{120} \Scadd{2}{3}{5}{6} + \ \nnu \\
& \quad &  \frac  {13}{180} \sca{1}{4}  +  \frac  {13}{360} \Scat{4}{2}{3} - \ \nnu \\
& \quad & \left.  \frac  {1}{18} \Scat{4}{5}{6} \right] \ \nnu \\
\kp1 \, \pm2 \, \pm3 \, \pm4 \, \pp5 \, \pp6 & \quad & \facs{1}{} \left[ -  \frac  {19}{270} \Scaq{1}{2}{3}{4} +
\right.
\ \nnu \\
& \quad &  \frac  {4}{135} ( \sca{2}{3} + \sca{2}{4} + \sca{3}{4} )  + \ \nnu \\
& \quad &  \frac  {13}{180} \Scat{1}{5}{6} + \ \nnu \\
& \quad &  \frac  {1}{180} \Scadt{5}{6}{2}{3}{4} - \ \nnu \\
& \quad & \left.  \frac  {1}{18} \sca{5}{6} \right] \ \nnu \\
\kp1 \, \pm2 \, \p03 \, \p04 \, \et5 \, \et6 & \quad & \facs{i}{} \left[ -  \frac  {1}{60} \sca{5}{6} +
 \frac  {1}{30} \Scat{1}{5}{6} +  \right. \ \nnu \\
& \quad &  \frac  {1}{15} \Scat{2}{5}{6} -  \frac  {13}{60} \sca{1}{2} - \ \nnu \\
& \quad &  \frac  {1}{120} \Scadd{3}{4}{5}{6} - \ \nnu \\
& \quad & \left.  \frac  {1}{24} \Scat{1}{3}{4} +  \frac  {3}{40} \Scat{2}{3}{4} \right] \ \nnu \\
\kp1 \, \pm2 \, \pm3 \, \pp4 \, \et5 \, \et6 & \quad & \facs{i}{} \left[ -  \frac  {1}{30} \sca{5}{6} +
 \frac  {1}{15} \Scat{1}{5}{6} + \right. \ \nnu \\
& \quad &  \frac  {7}{120} \Scadd{5}{6}{2}{3} - \ \nnu \\
& \quad &  \frac  {31}{120} \Scat{1}{2}{3} +  \frac  {3}{20} \sca{2}{3} - \ \nnu \\
& \quad &  \frac  {1}{60} \Scat{4}{5}{6} -  \frac  {1}{12} \sca{1}{4} + \ \nnu \\
& \quad & \left.  \frac  {3}{40} \Scat{4}{2}{3} \right] \ \nnu \\
\k01 \, \p02 \, \p03 \, \p04 \, \et5 \, \et6 & \quad & \facs{i}{\sqrt{2}} \left[  \frac  {1}{60} \sca{5}{6} -
 \frac  {1}{30} \Scat{1}{5}{6} - \right. \nnu \\
& \quad &  \frac  {1}{60} \Scadt{5}{6}{2}{3}{4} + \ \nnu \\
& \quad &  \frac  {1}{10} \Scaq{1}{2}{3}{4} - \ \nnu \\
& \quad & \left.  \frac  {1}{20} ( \sca{2}{3} + \sca{2}{4} + \sca{3}{4} )  \right] \ \nnu \\
\k01 \, \pm2 \, \pp3 \, \p04 \, \et5 \, \et6 & \quad & \facs{i}{\sqrt{2}} \left[  \frac  {1}{30} \sca{5}{6} -
 \frac  {1}{15} \Scat{1}{5}{6} + \right. \ \nnu \\
& \quad &  \frac  {1}{60} \Scat{2}{5}{6}  +  \frac  {1}{12} \sca{1}{2} - \ \nnu \\
& \quad &  \frac  {2}{15} \Scat{4}{5}{6} +  \frac  {13}{30} \sca{1}{4} - \ \nnu \\
& \quad &   \frac  {3}{20} \sca{2}{4} +  \frac  {1}{60} \Scat{3}{5}{6} + \ \nnu \\
& \quad & \left.  \frac  {1}{12} \sca{1}{3} -  \frac  {3}{20} \sca{3}{4} \right] \ \nnu \\
\k01 \, \k02 \, \bk3 \, \p04 \, \p05 \, \p06 & \quad & \facs{i}{\sqrt{2}} \left[
-  \frac  {1}{45} \Scaq{3}{4}{5}{6} + \right. \ \nnu \\
& \quad &  \frac{1}{30} \Scadt{1}{2}{4}{5}{6} - \ \nnu \\
& \quad & \left.  \frac  {2}{45} ( \sca{4}{5} + \sca{4}{6} + \sca{5}{6} ) \right] \ \nnu \\
\k01 \, \k02 \, \bk3 \, \pm4 \, \pp5 \, \p06 & \quad & \facs{i}{\sqrt{2}} \left[
 \frac  {13}{45} \sca{3}{4} -  \frac  {2}{45} \Scat{4}{1}{2} - \right. \ \nnu \\
& \quad &  \frac  {43}{90} \sca{3}{6} +  \frac  {16}{45} \Scat{6}{1}{2} + \ \nnu \\
& \quad &   \frac  {1}{30} \sca{4}{6} +  \frac  {1}{18} \sca{3}{5} - \ \nnu \\
& \quad &  \frac  {1}{9} \Scat{5}{1}{2} -  \frac  {1}{10} \sca{4}{5} - \ \nnu \\
& \quad & \left.  \frac  {1}{5} \sca{5}{6} \right] \ \nnu \\
\k01 \, \p02 \, \p03 \, \p04 \, \p05 \, \p06 & \quad & \facs{i}{\sqrt{2}} \left[
 \frac  {1}{300} \Scas{1}{2}{3}{4}{5}{6} - \right. \ \nnu \\
& \quad &  \frac  {1}{600} ( \sca{2}{3} + \sca{2}{4} + \sca{2}{5} + \ \nnu \\
& \quad & \sca{2}{6} + \sca{3}{4} + \sca{3}{5} + \sca{3}{6} +  \ \nnu \\
& \quad & \left. \sca{4}{5} + \sca{4}{6} + \sca{5}{6} ) \right] \ \nnu \\
\k01 \, \pm2 \, \pp3 \, \p04 \, \p05 \, \p06 & \quad & \facs{i}{\sqrt{2}} \left[
-  \frac  {13}{180} \sca{1}{2} +  \right. \ \nnu \\
& \quad &  \frac  {19}{270} \Scaq{1}{4}{5}{6} +  \ \nnu \\
& \quad &  \frac  {7}{540} \Scaq{2}{4}{5}{6} - \ \nnu \\
& \quad &  \frac  {13}{270} ( \sca{4}{5} + \sca{4}{6} + \sca{5}{6} ) - \ \nnu \\
& \quad & \left.  \frac  {13}{180} \sca{1}{3} +  \frac  {7}{540} \Scaq{3}{4}{5}{6} \right] \ \nnu \\
\k01 \, \pm2 \, \pm3 \, \pp4 \, \pp5 \, \p06 & \quad & \facs{i}{\sqrt{2}} \left[
-  \frac  {13}{180} \Scat{1}{2}{3} +  \frac  {1}{18} \sca{2}{3} + \right. \ \nnu \\
& \quad &  \frac  {16}{45} \sca{1}{6} -  \frac  {13}{180} \Scat{6}{2}{3} - \ \nnu \\
& \quad &  \frac  {13}{180} \Scat{1}{4}{5} + \ \nnu \\
& \quad &  \frac  {1}{36} \Scadd{2}{3}{4}{5} - \ \nnu \\
& \quad & \left.  \frac  {13}{180} \Scat{6}{4}{5} +  \frac  {1}{18} \sca{4}{5} \right] \ \nnu \\
\k01 \, \kp2 \, \km3 \, \p04 \, \p05 \, \p06 & \quad & \facs{i}{\sqrt{2}} \left[
-  \frac  {14}{45} \sca{1}{3} -  \frac  {13}{90} \sca{1}{2} +  \frac  {1}{6} \sca{2}{3} + \right. \ \nnu \\
& \quad &  \frac  {29}{270} \Scaq{1}{4}{5}{6} + \ \nnu \\
& \quad &  \frac  {13}{270} \Scaq{3}{4}{5}{6} - \ \nnu \\
& \quad &  \frac  {1}{135} \Scaq{2}{4}{5}{6} - \ \nnu \\
& \quad &  \left.  \frac  {7}{135} ( \sca{4}{5} + \sca{4}{6} + \sca{5}{6} ) \right]  \ \nnu \\
\k01 \, \kp2 \, \km3 \, \pm4 \, \pp5 \, \p06 & \quad & \facs{i}{\sqrt{2}} \left[
-  \frac  {28}{45} \sca{1}{3} -  \frac  {13}{45} \sca{1}{2} + \right. \ \nnu \\
& \quad &  \frac  {1}{3} \sca{2}{3} -  \frac  {13}{90} \sca{1}{4} -  \frac  {3}{10} \sca{3}{4} + \ \nnu \\
& \quad &   \frac  {1}{30} \sca{2}{4} +  \frac  {4}{5} \sca{1}{6} +  \frac  {19}{45} \sca{3}{6} - \ \nnu \\
& \quad &  \frac  {11}{45} \sca{2}{6} -  \frac  {29}{90} \sca{4}{6} -  \frac  {1}{90} \sca{1}{5} + \ \nnu \\
& \quad &  \frac  {1}{6} \sca{3}{5} +  \frac  {1}{6} \sca{2}{5} -  \frac  {2}{15} \sca{4}{5} + \ \nnu \\
& \quad & \left.  \frac  {13}{90} \sca{5}{6} \right] \ \nnu \\
\kp1 \, \k02 \, \bk3 \, \pm4 \, \p05 \, \p06 & \quad & \facs{i}{} \left[
-  \frac  {43}{180} \sca{2}{3} +  \frac  {43}{180} \sca{1}{3} + \right. \ \nnu \\
& \quad &  \frac  {16}{45} \sca{3}{4} +  \frac  {7}{60} \sca{2}{4} -  \frac  {73}{180} \sca{1}{4} - \ \nnu \\
& \quad &  \frac  {13}{90} \Scat{3}{5}{6} -  \frac{1}{45} \Scat{2}{5}{6} + \ \nnu \\
& \quad &  \frac  {1}{15} \Scat{1}{5}{6} +  \frac  {1}{10} \Scat{4}{5}{6} - \ \nnu \\
& \quad & \left.  \frac  {1}{15} \sca{5}{6} \right] \ \nnu \\
\kp1 \, \k02 \, \bk3 \, \pm4 \, \pm5 \, \pp6 & \quad & \facs{i}{} \left[
-  \frac  {43}{90} \sca{2}{3} +  \frac  {43}{90} \sca{1}{3} + \right. \ \nnu \\
& \quad &  \frac  {17}{180} \Scat{3}{4}{5} +  \frac  {11}{180} \Scat{2}{4}{5} - \ \nnu \\
& \quad &  \frac  {67}{180} \Scat{1}{4}{5} -  \frac  {1}{30} \sca{4}{5} - \ \nnu \\
& \quad &  \frac  {1}{18} \sca{3}{6}  +  \frac{1}{45} \sca{2}{6} + \ \nnu \\
& \quad & \left.  \frac  {1}{5} \sca{1}{6} +  \frac  {3}{20} \Scat{6}{4}{5} \right] \ \nnu \\
\eea
}
\end{document}